\documentstyle[sprocl]{article}
\begin{document}
\newcommand{\be}{\begin{eqnarray}}
\newcommand{\ee}{\end{eqnarray}}
\def\lo{\langle 0 |}
\def\ro{ | 0 \rangle }
\def\fc{ f_{\eta'}^{(c)} }
\def\gmf{\gamma _{5}}
\def\la{\langle }
\def\ra{ \rangle }
\def\el{ \langle \eta'| }
\def\er{ | \eta' \rangle}
\def\gmmu{\gamma _{\mu}}
\def\gmnu{\gamma_{\nu}}
\def\gmf{\gamma _{5}}
\newcommand{\beq}{\begin{equation}}
\newcommand{\eeq}{\end{equation}}
\newcommand{\bea}{\begin{eqnarray}}
\newcommand{\eea}{\end{eqnarray}}
\def\atop{ \frac{ \alpha_{s}}{4 \pi} G_{\mu \nu} \tilde{G}_{\mu \nu} }
\title{Intrinsic Strange/Charmed Quarks  
 Inside of a Strangeless/Charmless Hadron.  
 }
\author{Ariel~Zhitnitsky\footnote{\it Talk  
given  at   the
``HADRON 97''.
BNL, Brookhaven,  August 24-28, 1997.}
}
\address{
        Physics and Astronomy Department \\
        University of British Columbia \\
          Vancouver, BC V6T 1Z1, Canada  } 
  \maketitle\abstracts{   
We discuss few, apparently different, but actually tightly related problems:
a)Strangeness in the nucleon; b) 
      $ B \rightarrow \eta'X $ decays; c)
Intrinsic charm in  the proton spin problem.
 We   argue that all these problems have a common origin
and related to 
the OZI rule violation
in  $0^{\pm}$ vacuum channels . 
It leads to a noticeable role of a   nonvalence component
with $0^{\pm}$ quantum numbers  in a hadron.}

{\bf 1.}
Nowadays it is almost accepted that a {\bf nonvalence} component
in a hadron could be very high, much higher than naively
one could expect from the   perturbative estimations.
Experimentally, such a phenomenon was observed in a number
of places. Let me mention only one of them.
As is known,  the
matrix element $\la N|\bar{s}s|N\ra$ is
    not vanishing, as naively one could expect,     
 but rather, is the same order of magnitude as 
valence matrix element $\la N|\bar{d}d|N\ra$. 
  Standard arguments, see e.g.\cite{Donoghue1},
which lead to this conclusion are based on the
analysis of  the so-called $\sigma$ term:
$
\label{1a}
\frac{m_u+m_d}{2}\la p|\bar{u}u+\bar{d}d|p\ra=45 MeV.
$
Assuming octet-type $SU(3)$ breaking 
to be responsible for the mass splitting in the baryon octet, we find 
\be
\label{5}
\la p|\bar{u}u|p\ra\simeq  3.5,~
\la p|\bar{d}d|p\ra\simeq  2.8,~
\la p|\bar{s}s|p\ra\simeq  1.4.
\ee
  In spite of the very rough estimations presented above,
we believe that a reader is     convinced that :
{\bf a)} a magnitude of the nucleon matrix element 
for $\bar{s}s$ is not small;
{\bf b)} the large value  for this matrix element is due to the
nontrivial QCD vacuum structure where vacuum expectation values
of $u,d,s$ quarks are developed and they have   the same 
order in magnitude:
$\la p|\bar{s}s|p\ra\sim
\la 0|\bar{s}s |0\ra \sim \la 0|\bar{d}d |0\ra $.
This result is in a variance with the standard OZI rule expectation
  predicting that  any non-valence matrix element is suppressed
in comparison with a similar in structure, but valence one,
 see ref.\cite{arz}
for details on the QCD explanation of the OZI rule violation.
{\bf c)}The vacuum channels are very special
in a sense that  the OZI rule in those
channels  is badly broken  while in the vector channel the 
OZI rule works well.  
 
 {\bf 2.} Now   we want to  discuss a similar example
  where we believe   the OZI rule violation emerges like in 
eq. (\ref{5}).
Recently CLEO has reported\cite{CLEO} a very 
large branching ratio for the 
  production of $\eta'$ :
\beq
\label{Kim}
Br(B \rightarrow \eta'+X) \simeq 7.5
   \cdot 10^{-4}; ~~~
 Br(B \rightarrow \eta'+K)\simeq 
7.8
  \cdot 10^{-5}
\eeq
 To get a feeling of how large these numbers are, we 
present for comparison the  branching ratios for the 
inclusive and exclusive  productions of $J/\psi$ meson \cite{PDF}:
\beq
\label{psi}
Br(B \rightarrow J/\psi (direct)  +X ) \simeq  8.0
   \cdot 10^{-3}; ~~~~~
Br(B \rightarrow J/\psi  +K) \simeq  1
  \cdot 10^{-3}
\eeq
These processes are due to the Cabbibo favored 
$b\rightarrow\bar{c}cs$ decay which is largest possible 
amplitude without   charmed hadrons (like $D, D_s, \Lambda_c...$)
in the final state.
The comparison of these two numbers shows that
the amplitudes of processes (\ref{Kim})  
are only by a factor of $3$ less than the    Cabbibo favored 
amplitudes $b\rightarrow\bar{c}cs\rightarrow  J/\psi s$(\ref{psi}).
It is clear that   data (\ref{Kim}) is in severe contradiction 
with a standard view of the process at the 
quark level as a decay of the $b$-quark into 
 light quarks which could be naively suggested  
 keeping in mind the standard picture of   
 $ \eta' $ as  a SU(3) singlet
meson made of the $u-$, $d-$ and $s-$quarks. In this picture 
  decay (\ref{Kim}) must be proportional to the Cabbibo suppression
factor $V_{ub}$, and therefore the standard 
approach has no chance to explain 
  data (\ref{Kim}).
  Indeed, an assumption that the  $ \eta' $ 
is made exclusively of light quarks leads to the following ratio
for two pseudoscalar particles $\eta'$ and $\eta_c(1S)$ :
\bea
\label{1}
 \frac{\Gamma(B \rightarrow  \eta'+X)}{\Gamma(B \rightarrow 
 \eta_c(1S)+X)} \sim\frac{1}{3} (\frac{V_{bu}}{V_{bc}})^2\left(\frac{
f_{\eta'}}{ f_{\eta_c}} \right)^2   
\left(\frac{\Omega_{b \rightarrow  \eta'+X}}{
 \Omega_{b \rightarrow \eta_c+X}}\right)
\sim  3\cdot 10^{-4}.
\eea
Here $\Omega_{b \rightarrow  \eta'+X}$ and 
$\Omega_{b \rightarrow \eta_c+X}$
are the corresponding phase volumes for two inclusive decays;
$(\frac{V_{bu}}{V_{bc}})\simeq 0.08$. The matrix element 
$ \la \eta'(p) | 
\bar{u} \gmmu \gmf u \ro =\frac{-i}{\sqrt{3}}f_{\eta'}p_{\mu}
\simeq  (0.5\div 0.8)\frac{-i}{\sqrt{3}}f_{\pi}p_{\mu}$ is 
known numerically; $f_{\eta_c}\simeq 400 MeV$
can be estimated from the   $\eta_c\rightarrow\gamma\gamma $ decay. 
  Therefore, the standard mechanism
yields a very small contribution in comparison 
with   data (\ref{Kim}):
 $ 
Br(B \rightarrow  \eta'+X) \sim 1.5\cdot 10^{-6}
$. We should mention  that the factorization procedure used in 
the estimate (\ref{1}) does
not work well. A  phase factor 
introduced into this formula  is also a
rough simplification: in reality, an inclusive spectrum
is much more complicated function than 
a simple factor $ \Omega_{b \rightarrow  \eta'+X}$
obtained  as a result of two-particle decay of a colorful heavy quark
$b\rightarrow\eta'(\eta_c)+d(s)$ instead of the physical $B$ meson.
However, it is obvious that all these effects due to a 
non-factorizability, gluon corrections, 
  as well as  
$ O(1/m_b , 1/N) $ terms omitted in (\ref{1}),
cannot substantially change our estimate. 
We therefore conclude that the image of the $ \eta' $ meson 
as the
SU(3) singlet quark state made exclusively 
of the $ u,d,s $ quarks is not adequate 
to the problem at hand.
 
 {\bf 3.}  
   In view of the failure of the 
standard approach to the $ B \rightarrow \eta' +X$ decay 
which treats the  $ \eta' $  
as the
SU(3) singlet quark state made exclusively 
of the $ u,d,s $ quarks,   we suggest an 
alternative mechanism for the $ B \rightarrow\eta'+X $ decay 
which is specific to the uniqueness of the  
  $ \eta' $. It has been known \cite{Witten,Ven},
that the $ \eta' $ is {\bf  a messenger between two worlds}:
the world of light hadrons
and a less studied world  of gluonia.  In other words, it 
is a very special meson strongly coupled to gluons.  
We suggest the following picture for the process of interest:  
 the $ b \rightarrow c\bar{c} s $ 
decay is followed by the conversion of the $c\bar{c}$-pair into
 the $ \eta'$.
This means that the matrix element 
$ \lo \bar{c} \gmmu \gmf c | \eta'(p) \ra = i \fc p_{\mu}$
 is not zero
due to the $ c \bar{c}\rightarrow  gluons $ transition\cite{HZ}: 
 \beq 
\label{13}
\fc = - \frac{1}{16 \pi^2 m_{\eta'}^2 } \frac{1}{m_{c}^2}
\lo g^3 f^{abc} G_{\mu \nu}^a \tilde{G}_{\nu \alpha}^b 
G_{\alpha \mu}^c \er + 0(1/m_c^4)+...
\eeq
Of course,
since one deals here with virtual c-quarks, this matrix 
element is 
suppressed by the $ 1/m_{c}^2 $. However,
 the c-quark is not very heavy, and the suppression
$ 1/m_{c}^2 $ is not large numerically. At the same time,
 the Cabbibo
enhancement of the $ b \rightarrow c $ transition in
 comparison
to $ b \rightarrow u $ is a much more important factor 
which makes this mechanism work. One can estimate $\fc = ( 50 \sim 180) \; MeV $
  indirectly \cite{HZ}using a combination of the OPE, large
 $N$ approach and 
QCD low energy theorems. If one assumes the
 saturation of the experimental 
data (\ref{Kim})
by suggested mechanism one  obtains $\fc(exp.) \simeq 140 MeV$,
see\cite{HZ} for details.

{\bf 4.} Here we want to calculate $\fc $ directly 
  using
the Interacting Instanton Liquid Model (IILM),  
see \cite{SS_96} for a review.  
 The calculation is based on the numerical evaluation of the
following two-point Euclidean correlation functions
\be
\label{10}
K_{22}(x)=\lo T g^2 G_{\mu\nu}^a \tilde {G_{\mu\nu}^a}(x) ,~~ g^2 G_{\mu\nu}^a \tilde {G_{\mu\nu}^a} (0) \ro 
\ee
and similar for $K_{23}(x)$ and $K_{33}(x)$
where two-gluon 
operator $g^2 G_{\mu\nu}^a \tilde {G_{\mu\nu}^a}$ is replaced by the 
three- gluon operator $  g^3 f^{abc}G_{\mu\nu}^a \tilde{G}^b_{\nu\lambda}
G_{\lambda\mu}^c(x)$ once (for $K_{23}(x)$) or twice
(for $K_{33}(x)$) correspondingly. 
 The magnitude  $\fc$ can be obtained from
  the calculation of the correlation functions:
\be
\label{15}
|\frac{\fc\sqrt{3}m_c^2}{f_{\eta'}}|=|\frac{K_{23}(x\rightarrow\infty)}
{K_{22}(x\rightarrow\infty)}|=
\sqrt{\frac{K_{33}(x\rightarrow\infty)}
{ K_{22}(x\rightarrow\infty)}}
\ee
The corresponding   measurements of $K_{23},K_{33}$
 both ratios entering (\ref{15})  
has been carrying out in ref. \cite{SZ} where it was  found 
the stabilization  at large enough x$>0.8 fm$ at the $same$ numerical value.
We take it as an indication that  $\eta'$  contribution
does in fact dominate in this region.
Final result of this calculations can be presented in the form\cite{SZ}:
 \be
 \label{result}
| \fc/f_{\eta'}|\simeq (0.85\sim 1.22 ).
\ee
The obtained result is in a fair agreement with  ``experimental'' value
$\fc(exp.)\simeq 140 MeV$
needed to explain CLEO measurements, inside the uncertainties.
   
{\bf 5.} The next logical question to ask is whether
the connection between strong instanton fields and charm lead to 
phenomena unrelated to $\eta'$. One intriguing direction 
is to study the   ``intrinsic charm'' (see
 e.g. \cite{brodsky}) of other hadrons.
In particular, one could  consider the  charm contribution 
to the
 spin  of the nucleon\cite{3}. The relevant
 matrix element is  the axial current of the charmed quark, 
$\la N  | \bar{c} \gmmu \gmf c | N \ra = 
g_A^{(c)}  \bar{N}      \gmmu \gmf N 
$.
It
could be generated e.g. by the $ \eta' $ ``cloud'' of the nucleon. 
Assuming now the  $ \eta' $ dominance in this matrix 
element one could get the following 
   Goldberger-Treiman type relation\cite{3}
$
g_A^{(c)} = \frac{1}{2 M_{N}} g_{\eta' NN} \fc
$.
Although the  value
of $ g_{\eta' NN } $ is unknown, and its
phenomenological estimates 
 vary significantly
$ g_{\eta' NN } = 3-7 $, one gets from this estimate a
surprisingly
large contribution   
$
 \la N  | \bar{c} \gmmu \gmf c | N \ra = (0..2\sim 0.5)
\bar{N}     \gmmu \gmf N   
$
 comparable to the light
 quark one, see\cite{KZ},\cite{3}! Ultimately, 
the contribution of the charmed quarks 
in polarized deep-inelastic scattering may be  tested
experimentally, by tagging the charmed quark jets (e.g. by COMPASS
experiment at CERN).

{\bf 6.}  The situation reminds me the $ J /\psi $ discovery in 1974, when
a charmonium state (``hidden charm") was observed simultaneously
in $ e^{+} e^{-} $ collisions at SLAC   and at the 
proton machine at Brookhaven. I believe that we 
are now facing a similar case, when different experimental groups
see the ``intrinsic charm" in polarized DIS   and in 
B-decays (\ref{Kim}) simultaneously, see\cite{3} for details.

\end{document}